\documentclass[aps,twocolumn,pre]{revtex4}
\usepackage[utf8]{inputenc}
\usepackage{graphicx}
\usepackage{latexsym}
\usepackage{amsmath}
\usepackage{amssymb}
\usepackage{psfrag}
\usepackage{pifont}
\usepackage{color}
\usepackage{subeqnarray}
\usepackage{lipsum}
\usepackage{bm}
\newcommand{\beq}{\begin{equation}}
\newcommand{\eeq}{\end{equation}}
\newcommand{\ba}{\begin{array}}
\newcommand{\ea}{\end{array}}
\newcommand{\bee}{\begin{eqnarray}}
\newcommand{\ec}{\end{center}}

\newcommand{\eee}{\end{eqnarray}}
\newcommand{\bc}{\begin{center}}

\makeatletter
\begin{document}

\title{
Spectral analysis for elastica 3-dimensional dynamics in a shear flow} 
\author{Lujia Liu}
\author{Pawe\l{} Sznajder}
\author{Maria L. Ekiel-Je\.zewska\footnote{Corresponding author. Email: mekiel@ippt.pan.pl}}
\affiliation{Institute of Fundamental Technological Research, Polish Academy of Sciences, Pawi\'nskiego 5b, 02-106 Warsaw, Poland}

\begin{abstract}
We present the spectral analysis of three-dimensional 
dynamics of an elastic filament in a shear flow of a viscous fluid at a low Reynolds number in the absence of Brownian motion. The elastica model is used. The fiber initially is almost straight at an arbitrary orientation, with small perpendicular perturbations in the shear plane and out-of-plane. To analyze the stability of both perturbations, equations for the eigenvalues and eigenfunctions 
are derived and solved  by the Chebyshev spectral collocation method. It is  shown that 
their crucial features 
are the same as in 
case of the two-dimensional elastica dynamics in shear flow [Becker and Shelley, Phys. Rev. Lett. 2001] and the three-dimensional elastica dynamics in the compressional flow [Chakrabarti et al., Nat. Phys., 2020]. We find a similar  dependence of 
the buckled shapes 
on the ratio of bending to hydrodynamic forces as in the simulations for elastic fibers of a nonzero thickness 
[Slowicka et al., New J. Phys., 2022].   
\end{abstract}

\date{\today}
\maketitle

\section{Introduction}
Dynamics of flexible micro and nano-objects in fluid flows have been recently widely investigated theoretically, numerically, and experimentally \cite{Linder2015,du2019dynamics,diamant2020}. The interest is motivated by potential applications to microorganisms  (e.g., diatoms \cite{Nguyen2014}, bacteria \cite{kantsler2012}, cells \cite{sinha2015dynamics} or actin \cite{harasim2013,liu2018}), and to artificially produced micro and nanofibers \cite{nunes2013,nakielski2015hydrogel}.  

For elastic filaments,   typical  time-dependent shape deformations and orientations have been analyzed for different values of the bending stiffness and aspect ratio. Buckling of elastic filaments has been studied in various fluid flows: extensional \cite{kantsler2012,manikantan2015,chakrabarti2020}, cellular \cite{young2007stretch,wandersman2010,quennouz2015,yang2017dynamics}, stagnation point \cite{guglielmini2012}, corner \cite{autrusson2011}, shear \cite{Forgacs1959a,forgacs1959particle,Becker2001}, and unidirectional \cite{Kurzthaler2023}. 

Many articles have focused on flexible filaments in  
shear flows \cite{harasim2013,Nguyen2014,sinha2015dynamics,Slowicka2015,liu2018,lagrone2019complex,kanchan2020numerical,Slowicka2020,zuk2021universal,xue2022shear,slowicka2022,bonacci2023dynamics}. The buckling instability of slender fibers located in the shear plane has been derived from the 2-dimensional spectral analysis of the elastica \cite{Becker2001}. The eigenvalues have been evaluated for a wide range of values of the elastoviscous number \cite{Becker2001}.

The goal of this work is to analyze the stability of elastica in a shear flow by solving a 3-dimensional spectral problem: not only for in-plane but also for  out-of-plane perturbations. Using the Chebyshev collocation method \cite{Trefethen2000}, we evaluate the eigenvalues and eigenfunctions and discuss the results.

\section{System and theoretical model} 
\subsection{Elastica 3-dimensional  equation}
A slender elastic 
filament 
is immersed in a shear flow  $\tilde{\bm{U}}\!=\!(\Dot{\gamma}y,0,0)$ of a fluid with the dynamic viscosity $\mu$. The ratio of the radius $R$ of the particle cross-section to its length $L$ 
is much smaller than unity, i.e., $R/L \ll~1$. The filament is assumed to be inextensible, and its 
length $L$ is used as the unit of length. 
The dimensionless position of the filament centerline is denoted as 
$\bm{x}(s,t)$, where $s\in [-1/2,1/2]$ is the arclength coordinate of 
a filament segment and $t$ is 
time. Owing to the 
filament inexensibility, 
$\bm{x}_s \cdot \bm{x}_s=1$. 
Here we ignore gravitational effects, Brownian forces, and  fluid inertia, assuming that the Reynolds number is much smaller than unity; only elastic and viscous forces are considered to result in the filament dynamics following from 
the Stokes equations. The slender body approximation \cite{batchelor1970slender} is used. 
With the length unit $L$, the time unit $\Dot{\gamma}^{-1}$ and $\bm{U}=\tilde{\bm{U}}/(\Dot{\gamma}L)$, the dimensionless elastica 3D governing equation is \cite{Audoly2000,Becker2001,guglielmini2012,Linder2015,Audoly2015}, 
\begin{equation}
    \eta (2 \bm{I}-\bm{x}_s\bm{x}_s)\cdot (\bm{x}_t-\bm{U}(\bm{x}))=(T(s,t)\bm{x}_s)_s-\bm{x}_{ssss},
    \label{eq:1}
\end{equation}
where $\eta=\frac{2\pi \mu \Dot{\gamma}L^4}{EI \ln(L/R)}$ denotes the ratio of the viscous drag to the elastic restoring forces \cite{guglielmini2012},  
$E$ is the Young's modulus of the filament and $I=\pi R^4/4$ is the area moment of inertia.
Here $T(s,t)$ represents the tension in the filament at point $s$ and time $t$.
\begin{figure}[b!]\vspace{-.4cm}
	\centerline{\includegraphics[width=0.5\textwidth]{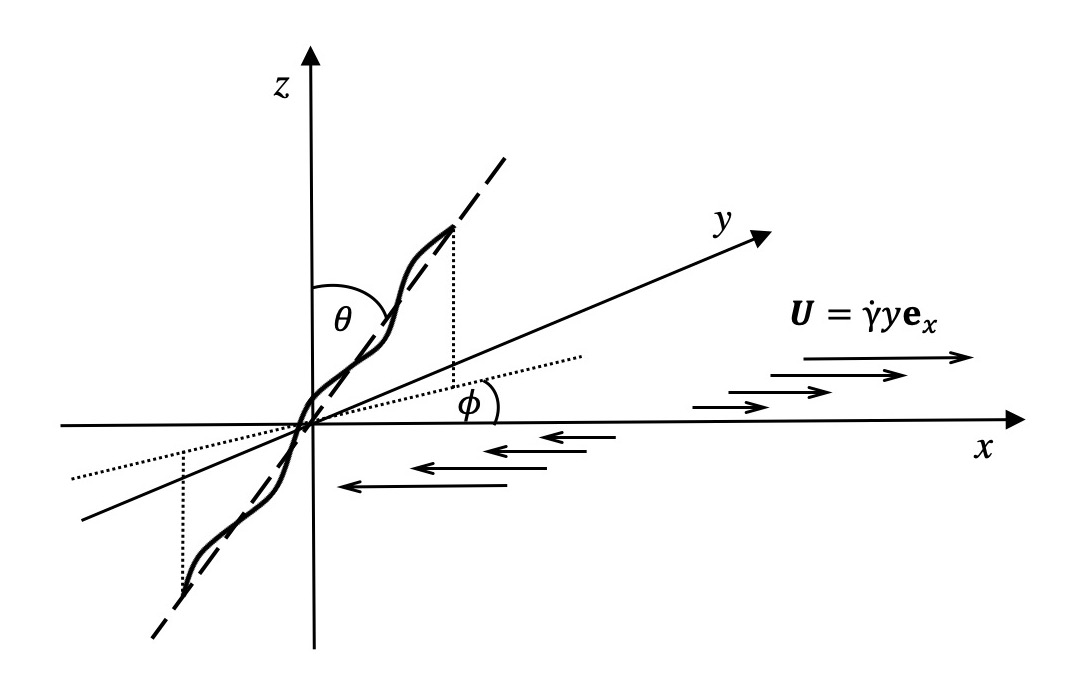}} \vspace{-.3cm}
	\caption{A schematic of a slender particle in a shear flow.}
	\label{fig:1}
\end{figure}

The particle is slightly deformed from a  straight shape at a certain orientation, 
described by the azimuthal and polar angles, $\phi (t)$ and $\theta (t)$, as   shown 
in Fig.~\ref{fig:1}. The azimuthal angle deals with the projection onto the shear plane $(x,y)$. Owing to symmetries, we can restrict to $0 \le \phi \le \pi$ and $0\le\theta \le \pi/2$ without loss of generality. 

The position vector $\bm{x}(s,t)$ can be described by 
\begin{equation}
    \bm{x}=s \bm{\lambda}+u \bm{\lambda}_{ i}+v \bm{\lambda}_{ o}, 
\end{equation}
where $u$ and $v$ are respectively the in-plane and out-of-plane deflections relative to a straight rod with the orientation determined by the unit vector $\bm{\lambda}=(\cos\phi \sin \theta, \sin\phi \sin\theta, \cos\theta)$. 
Here
$\bm{\lambda}_{ i}=(-\sin\phi, \cos\phi, 0)$ denotes the in-plane ($x, y$) unit vector perpendicular to $\bm{\lambda}$ and $\bm{\lambda}_{o}=(-\cos\phi\cos\theta,-\sin\phi\cos\theta, \sin\theta)$ denotes the out-of-plane unit vector perpendicular to both $\bm{\lambda}$ and $\bm{\lambda}_{i}$.

\subsection{Elastica equations for 
small 3D perturbations}

We assume that 
both perturbations, $u$, and $v$, are small and obtain a set of linearized equations. We present the zero-order and first-order equations, and then we consider the spectral problem that gives information about the stability of the perturbations. 

\begin{figure}[ht]
\vspace{-1.8cm} \centerline{\hspace{.04cm}
	\includegraphics[width=0.563\textwidth]{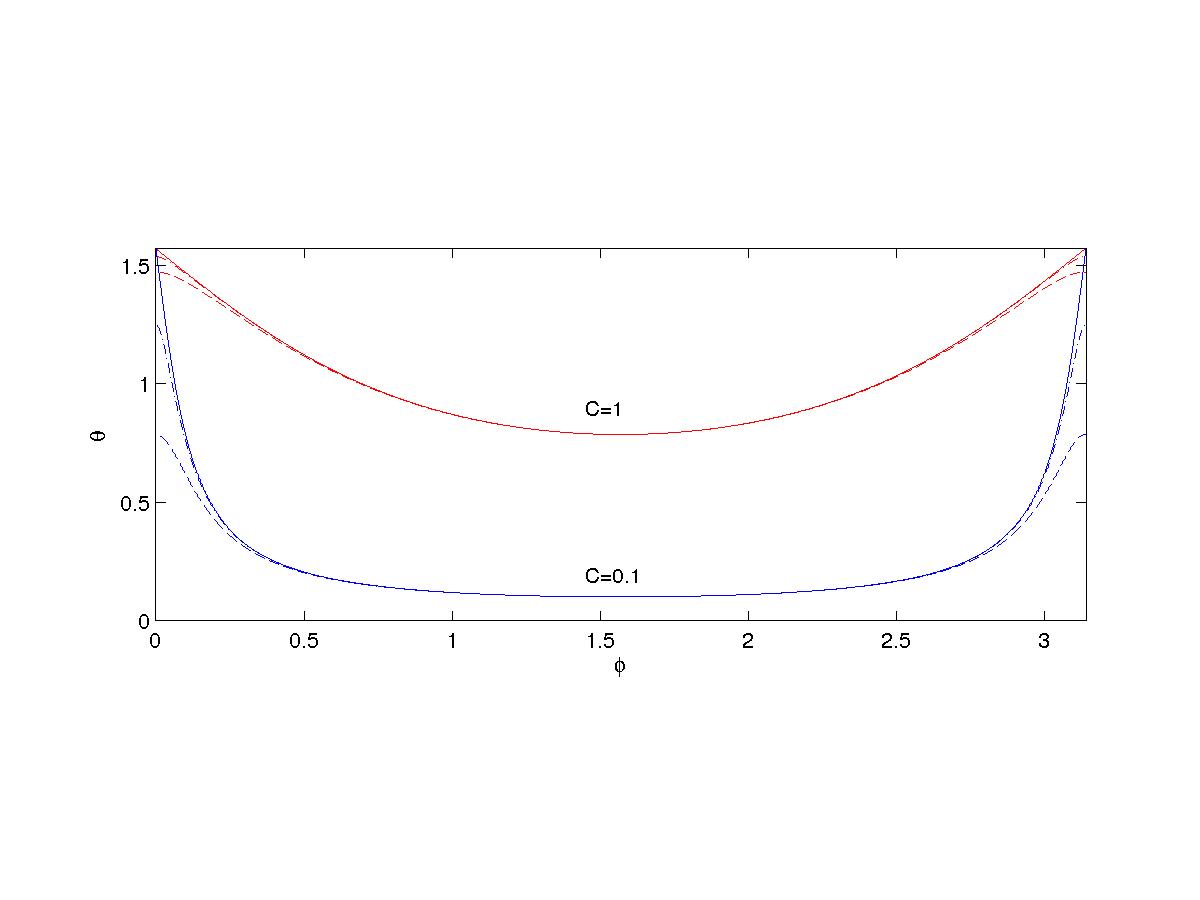}}\vspace{-1.8cm}
	\caption{
	Jeffery orbits $\theta(\phi)$ in a shear flow for spheroids with the aspect ratios $\ell$=10 (dashed line) and $\ell$=30 (dash-dotted line) 
	\cite{graham2018microhydrodynamics} 
	and for the zero-order elastica, Eq. 
	\eqref{eq:8} (solid line). The corresponding values of $C$ as indicated. 
	}
	\label{fig:5}
\end{figure}

Expanding the governing 
elastica equation (\ref{eq:1}), 
one 
obtains the following 
zero-order equations,
\bee
\Dot{\phi}&=&-\sin^2 \phi,
    \label{eq:6}\\
    \Dot{\theta}&=&
    \frac{1}{4}\sin(2\phi) \sin(2\theta),\label{eq:6a}\\
       T&=&-\frac{\eta}{4}\sin^2 \theta \sin(2\phi)  (s^2-\frac{1}{4}),
    \label{eq:5}
    \eee
  with the assumed boundary condition $T=0$ at the ends of the filament, i.e., for $s=\pm \frac{1}{2}$.
By solving Eqs \eqref{eq:6}-\eqref{eq:6a}, one 
obtains evolution of the filament orientation, 
\bee
    \cot\phi(t)&=&t+B,
    \label{eq:7}\\
    \tan\theta(t)&=& \frac{C}{\sin\phi(t)},
    \label{eq:8}
\eee
where $B$ and $C$ are constants dependent on initial conditions. 
The zero-order motion of elastica corresponds to the Jeffery orbit \cite{graham2018microhydrodynamics} in the limit of an infinitely thin rigid rod (the aspect ratio $\ell \rightarrow \infty$). In this limit, the time to approach (or leave) $\phi=0$ or $\pi$ is infinite, and therefore an elastica does not tumble, and as such, it can approximate dynamics of a fiber with a non-zero thickness only for the angle $\phi$ not very close to zero or $\pi$~
\cite{Slowicka2020,slowicka2022}. However, Jeffery trajectories $\theta(\phi)$ of the elastica zero-order motion and of a rigid rod with a sufficiently large aspect ratio, e.g.  $\ell \gtrsim 100$, are very close to each other, as illustrated in Fig. \ref{fig:5}.

Further, the first-order equations are 
derived from Eq.~\eqref{eq:1}. The in-plane perturbation $u$ satisfies the following equation, 
 \bee 
    &&\hspace{-0.6cm}
    u_{ssss}+2\eta u_t+ \eta \sin(2\phi) u+\eta \sin(2\phi) \sin^2 \theta s u_s\nonumber \\
    &&\hspace{-0.6cm}
    +\frac{\eta}{4} \sin(2\phi)\sin^2 \theta (s^2\! -\!\frac{1}{4})u_{ss}=0. 
    \label{eq:9}
    \eee
The equation for the out-of-plane perturbation $v$ has the form
    \bee
    &&\hspace{-0.6cm}
    v_{ssss}+2\eta v_t-\eta\sin(2\phi)\cos^2 \theta v  +2\eta \cos(2\phi)\cos(\theta) u
    \nonumber\\
    &&\hspace{-0.6cm} 
    +\eta \sin(2\phi) \sin^2 \theta s v_s
    +\frac{\eta}{4} \sin(2\phi)\sin^2 \theta (s^2\! -\!\frac{1}{4})v_{ss}=0. \nonumber \\
    \label{eq:10}
\eee

In the considered system, there are no external torques 
exerted on the filament ends,  
\beq u_{ss}=v_{ss}=0 \mbox{ at } s=\pm \frac{1}{2},\label{bc1}
\eeq
and external forces on the filament ends vanish, 
\beq u_{sss}=v_{sss}=0 \mbox{ at } s=\pm \frac{1}{2}.\label{bc2}
\eeq

\subsection{Eigenproblem and stability}
To estimate stability of the in-plane ($u$) and out-of-plane ($v$) perturbations, we consider the following eigenproblem,  
which is a 3D generalization of the in-plane study \cite{Becker2001}, and in the next section will be solved by the Chebyshev spectral collocation method \cite{Trefethen2000,Boyd2001}, also applied in \cite{Becker2001,chakrabarti2020}. 
We follow the standard assumption that
\begin{equation}
\{u, v\}=\{\Phi_u(s)\exp(\sigma t), \Phi_v(s)\exp(\sigma t)\}, 
\label{eks}
\end{equation}
where $\Phi_u$ and $\Phi_v$ are shapes of the in- and out-of-plane perturbations and $\sigma$ is the complex growth rate. 
\begin{figure*}[t!]
	\includegraphics[width=0.594\textwidth,trim={3.2cm 0cm 4.2cm 0.2cm},clip]{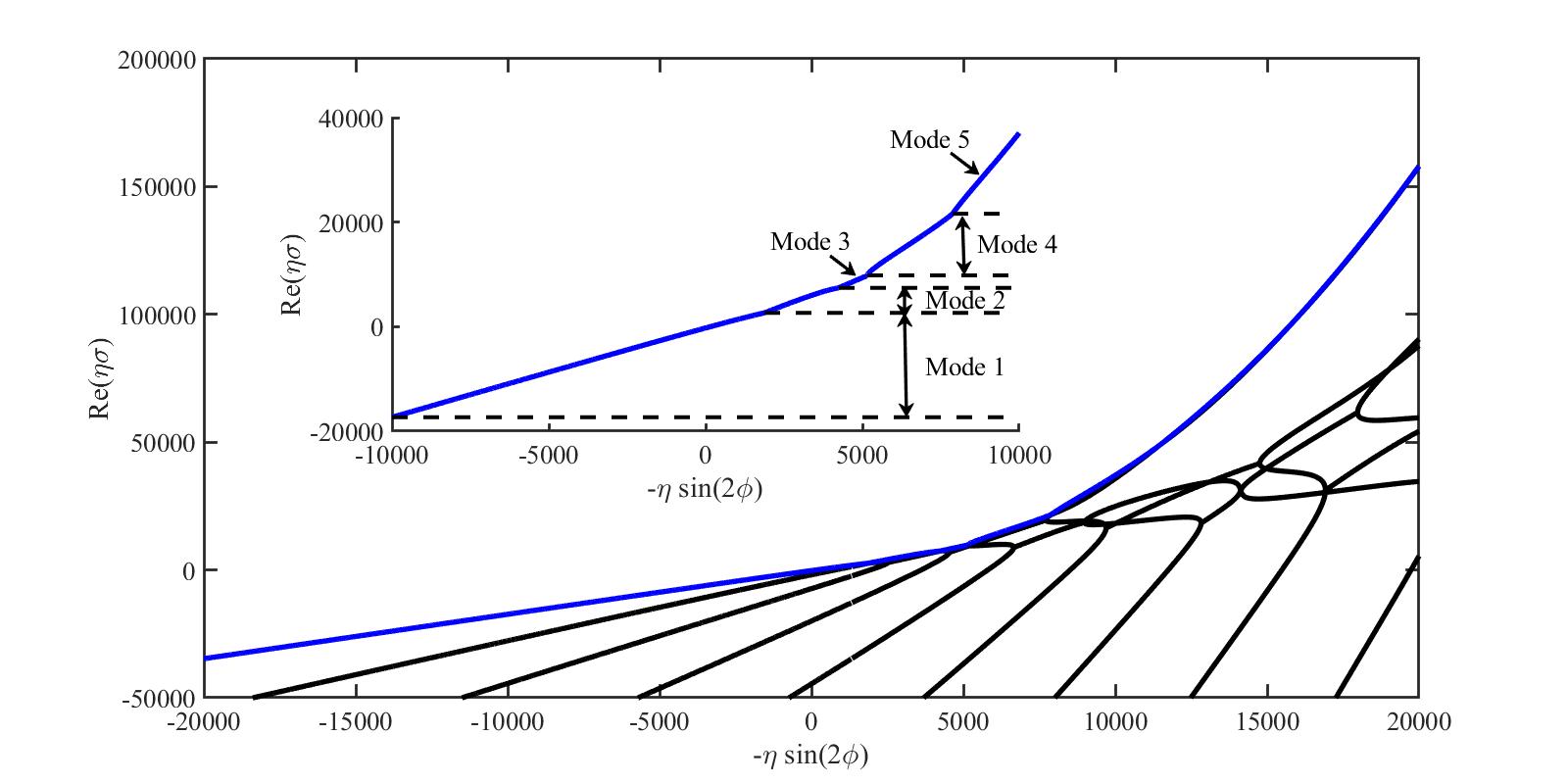}
	\includegraphics[width=0.40\textwidth,trim={3.34cm 0cm 20.15cm 0.2cm},clip]{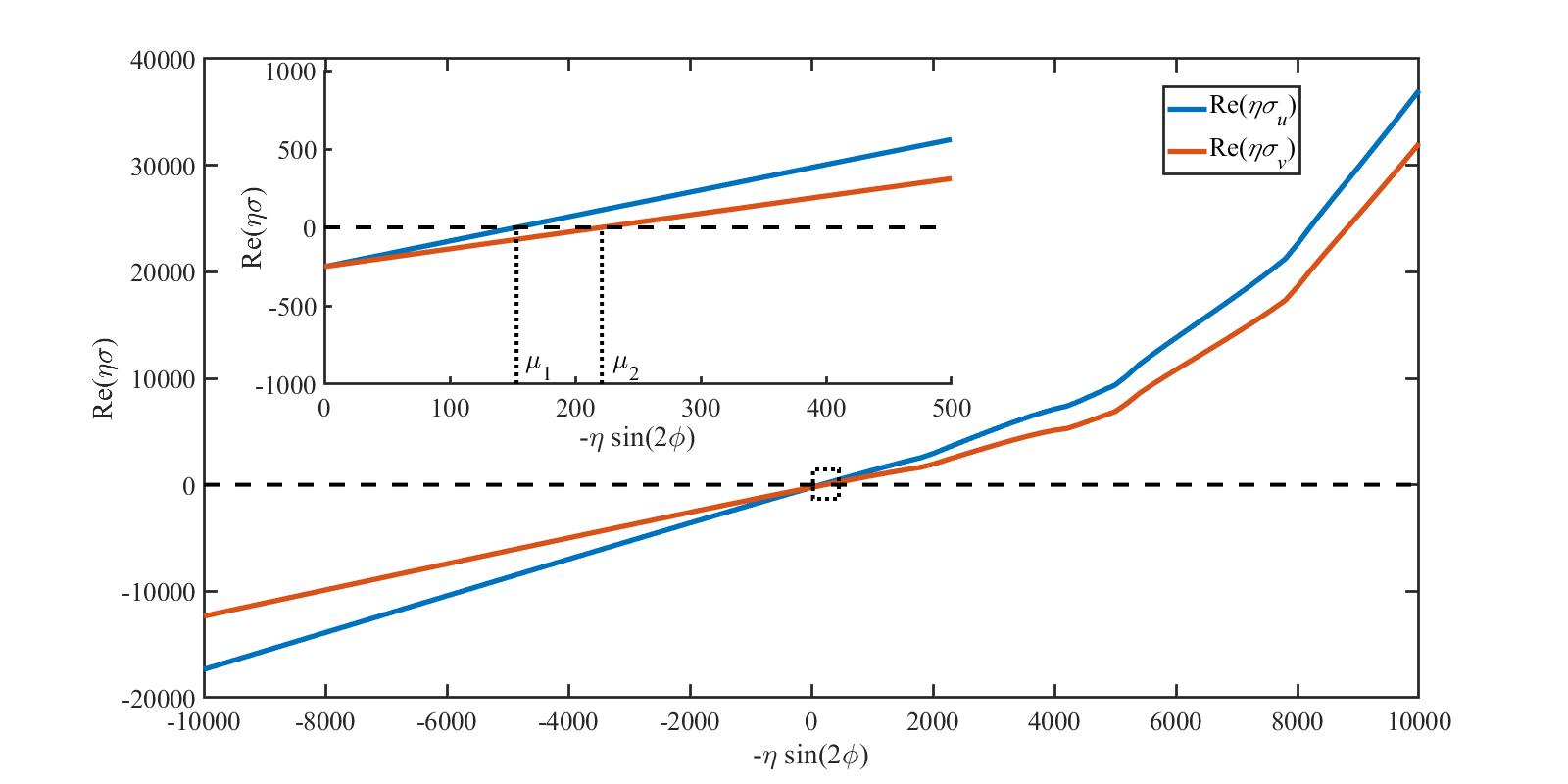}\vspace{-.1cm}
	\caption{Left: The real part of the  eigenspectrum $\eta \sigma\equiv\eta \sigma_u$, defined in 
	Eqs \eqref{eq:11p} and \eqref{Lplanar}, as a function of  $-\eta \sin(2\phi)$. Right: The largest eigenvalues, $\eta \sigma_u$ and $\eta \sigma_v$,  for the in-plane (blue) and out-of-plane (red) perturbations, $\Phi_u$ and $\Phi_v$, 
	respectively. 
}
	\label{fig:2}\label{fig:7}
	\vspace{.4cm}
	\includegraphics[width=0.51\textwidth,trim={3.73cm 0cm 4.2cm 0.2cm},clip]{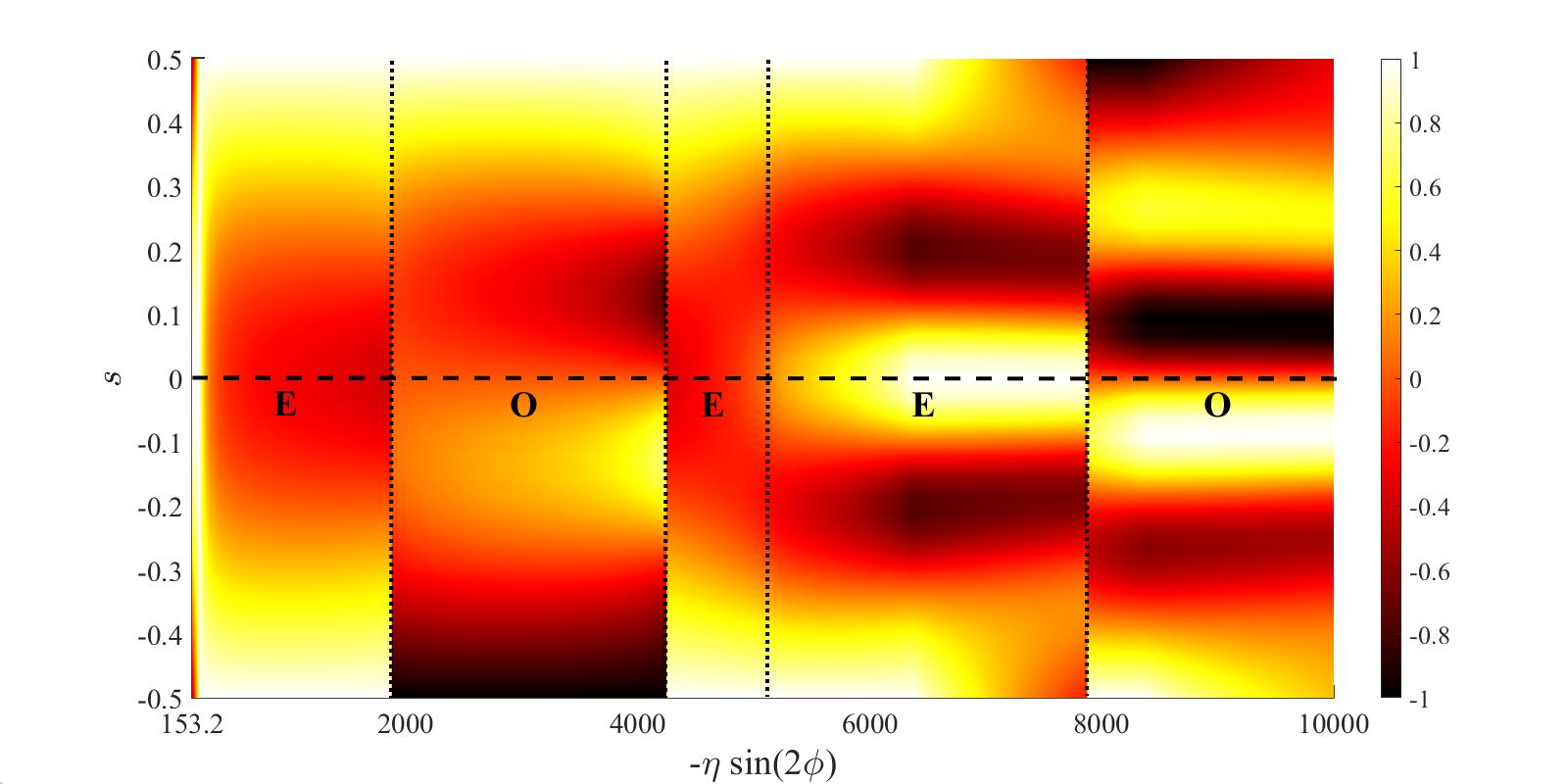}
\includegraphics[width=0.484\textwidth,trim={3.73cm 0cm 7.15cm 0.2cm},clip]{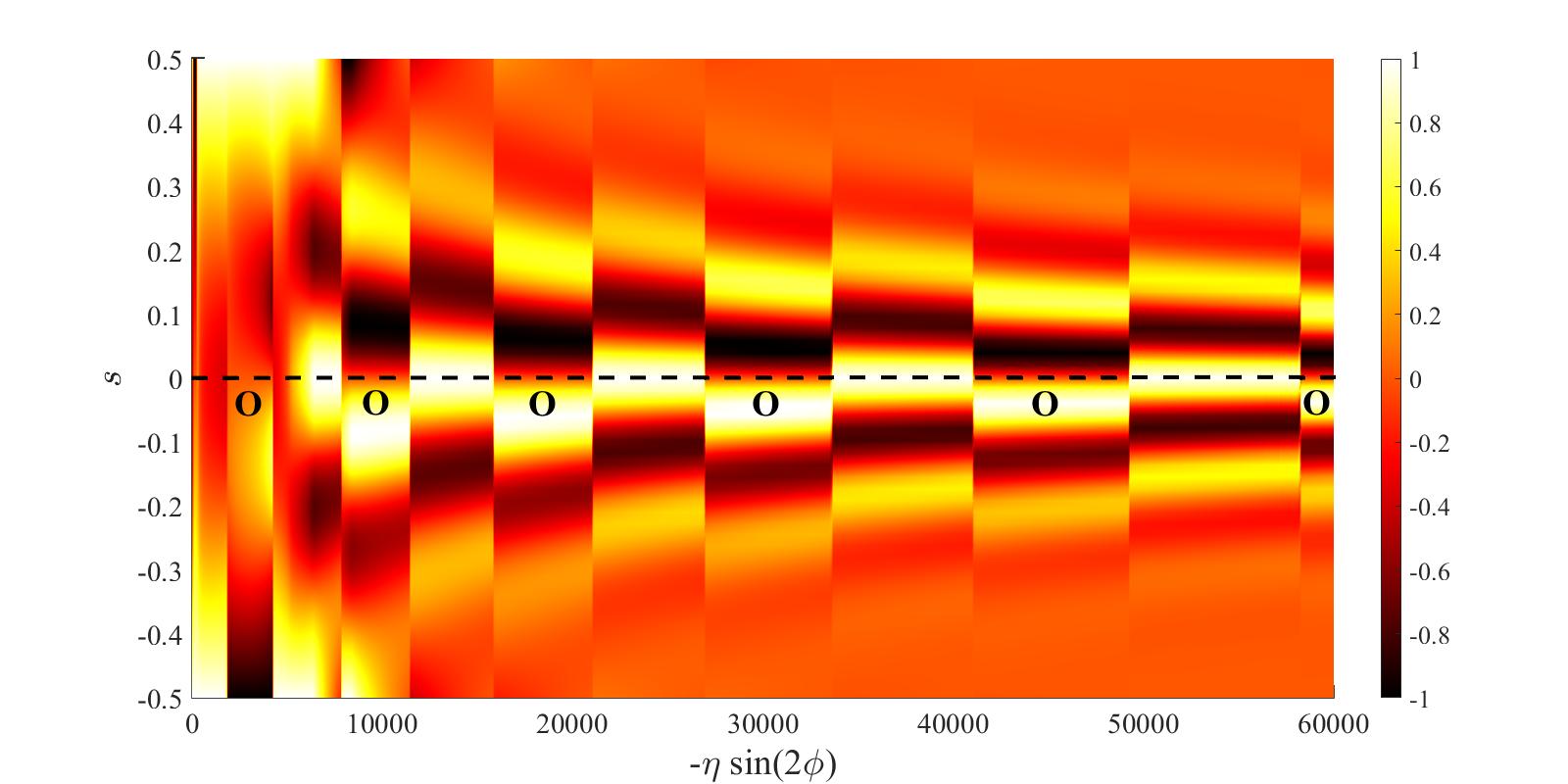} 	\vspace{-.65cm}
	\caption{
	The most unstable eigenfunctions $\Phi_u(s)$ are even (E) or odd (O), depending on  $-\eta \sin(2\phi)$. 
	Colors indicate values of $-1\le \Phi_u(s)\le 1$. 
	Left: 
	the most unstable modes 1-5 in the lowest range of 
	$-\eta \sin(2\phi)$, discussed in Ref. \cite{Becker2001}. Right: overview. 
	}
	\label{fig:12}
\end{figure*} 
Then we obtain the following spectral problem,
\bee 
  &&  \hspace{-0.6cm} 2\eta (\sigma+\frac{\sin(2\phi)}{2})\Phi_u=\mathcal{L}[\Phi_u],
    \label{eq:11}
\\ 
 && \hspace{-0.6cm}  2\eta (\sigma-\frac{\sin(2\phi)}{2}\cos^2 \theta) \Phi_v=\mathcal{L}[\Phi_v]-2\eta \cos(2\phi)\cos\theta \Phi_u, \nonumber \\
    \label{eq:12}
\eee
where $\mathcal{L}$ denotes the differential operator
\begin{equation}
    \mathcal{L}\!=\!-\frac{\partial^4}{\partial s^4}-\frac{\eta}{4}(s^2-\frac{1}{4})\sin(2\phi)\sin^2 \!\theta \frac{\partial^2}{\partial s^2}-\eta s \sin(2\phi)\sin^2 \!\theta \frac{\partial}{\partial s},\label{L}
\end{equation}
and the boundary conditions follow from Eqs \eqref{bc1}-\eqref{bc2}.

\section{Spectrum and eigenfunctions}
We 
consider the eigenproblem for the in-plane and out-of-plane perturbations when $\theta=\pi/2$. The spectral problem for the in-plane perturbations at $\theta=\pi/2$ was solved by Becker \& Shelley \cite{Becker2001}, who were interested in this special case because for an almost straight elastic filament performing a Jeffery orbit 
in the plane of shear, i.e. at $\theta=\pi/2$, the hydrodynamic forces exerted on it by the fluid 
are the largest. 

Here we extend the results from Ref. \cite{Becker2001} by taking into account not only in-plane, but also out-of-plane perturbations, $\Phi_u$ and $\Phi_v$, 
respectively.
In case of $\theta=\pi/2$, the first-order Eqs \eqref{eq:11}-\eqref{L} for the exponentially growing perturbations \eqref{eks} have the form,
\bee
  &&  (2\eta \sigma_u+\eta \sin(2\phi))\Phi_u=\tilde{\mathcal{L}}[\Phi_u],
    \label{eq:11p}
\\
 &&   2\eta \sigma_v \Phi_v=\tilde{\mathcal{L}}[\Phi_v],
    \label{eq:12p}
\eee
with
\begin{equation}
    \tilde{\mathcal{L}}=-\frac{\partial^4}{\partial s^4}-\frac{\eta\sin(2\phi)}{4}(s^2-\frac{1}{4}) \frac{\partial^2}{\partial s^2}-\eta  \sin(2\phi)s \frac{\partial}{\partial s}.\label{Lplanar}
\end{equation}

Eqs \eqref{eq:11p} and \eqref{eq:12p} for the in-plane and out-of-plane perturbations 
decouple from each other. The eigenfunction $\Phi_u$ belonging to an eigenvalue $\sigma_u$ is identical to the eigenfunction $\Phi_v$ belonging to the shifted eigenvalue $\sigma_u + \sin(2\phi)/2$. Therefore, 
we 
focus on the solutions to Eqs  \eqref{eq:11p} and \eqref{Lplanar}. The generalization for the  out-of-plane perturbations is straightforward.

We solve the eigenvalue problem for $\Phi_u$, defined by Eqs~\eqref{eq:11p} and \eqref{Lplanar}, 
by the Chebyshev spectral collocation method.
The calculated 
eigenspectrum for the in-plane perturbation $\Phi_u$ is demonstrated in the left panel of Fig.~\ref{fig:2}. 
It agrees well with the result presented in Fig.~2 of Ref. 
\cite{Becker2001}; 
also, the ranges of $-\eta \sin(2\phi)$ corresponding to 
modes 1-5 
are the same. 
\begin{figure*}
\includegraphics[width=0.495\textwidth,trim={5.99cm 0cm 4.35cm 0.2cm},clip]{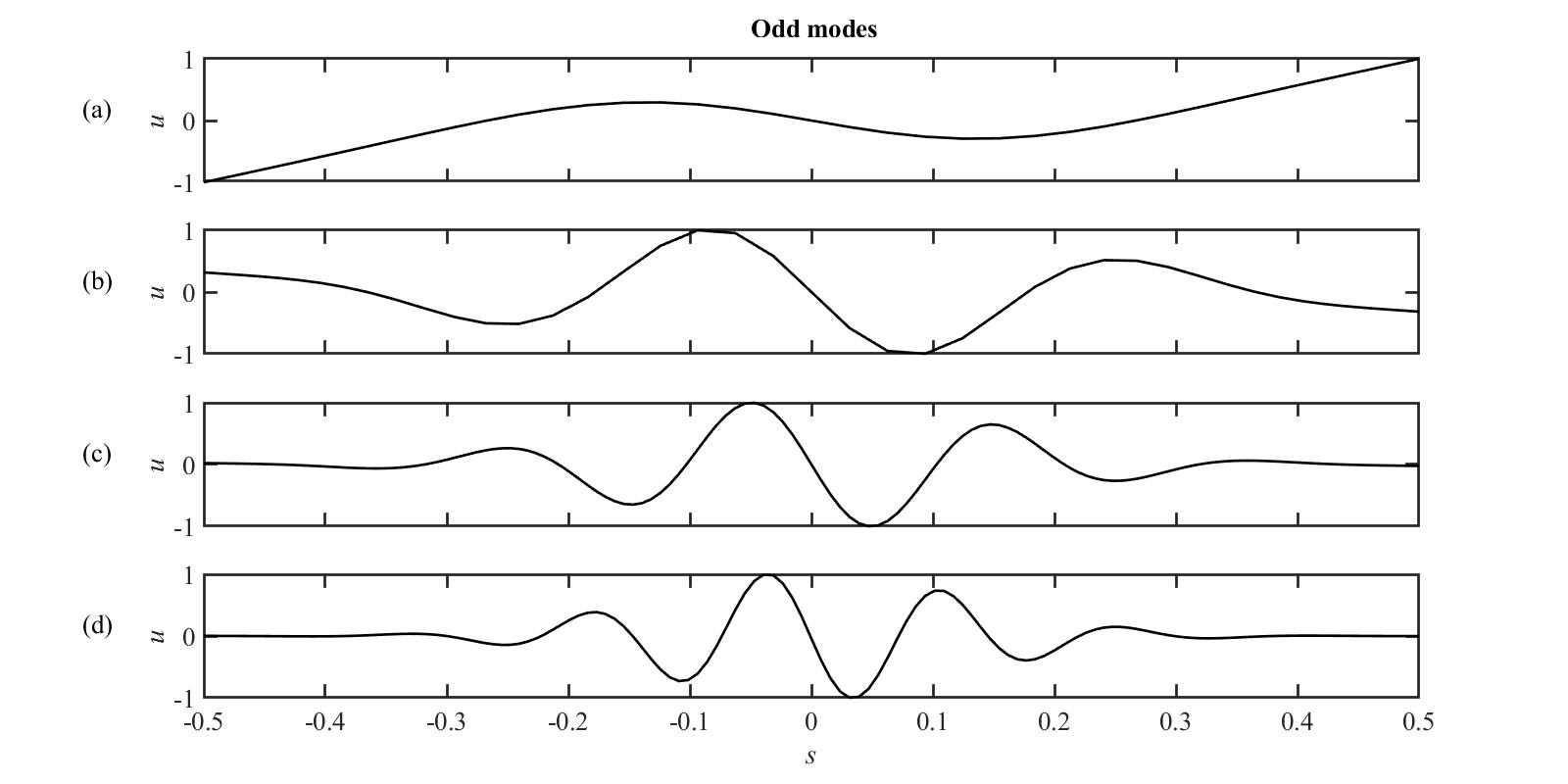}
	\includegraphics[width=0.495\textwidth,trim={5.99cm 0cm 4.35cm 0.2cm},clip]{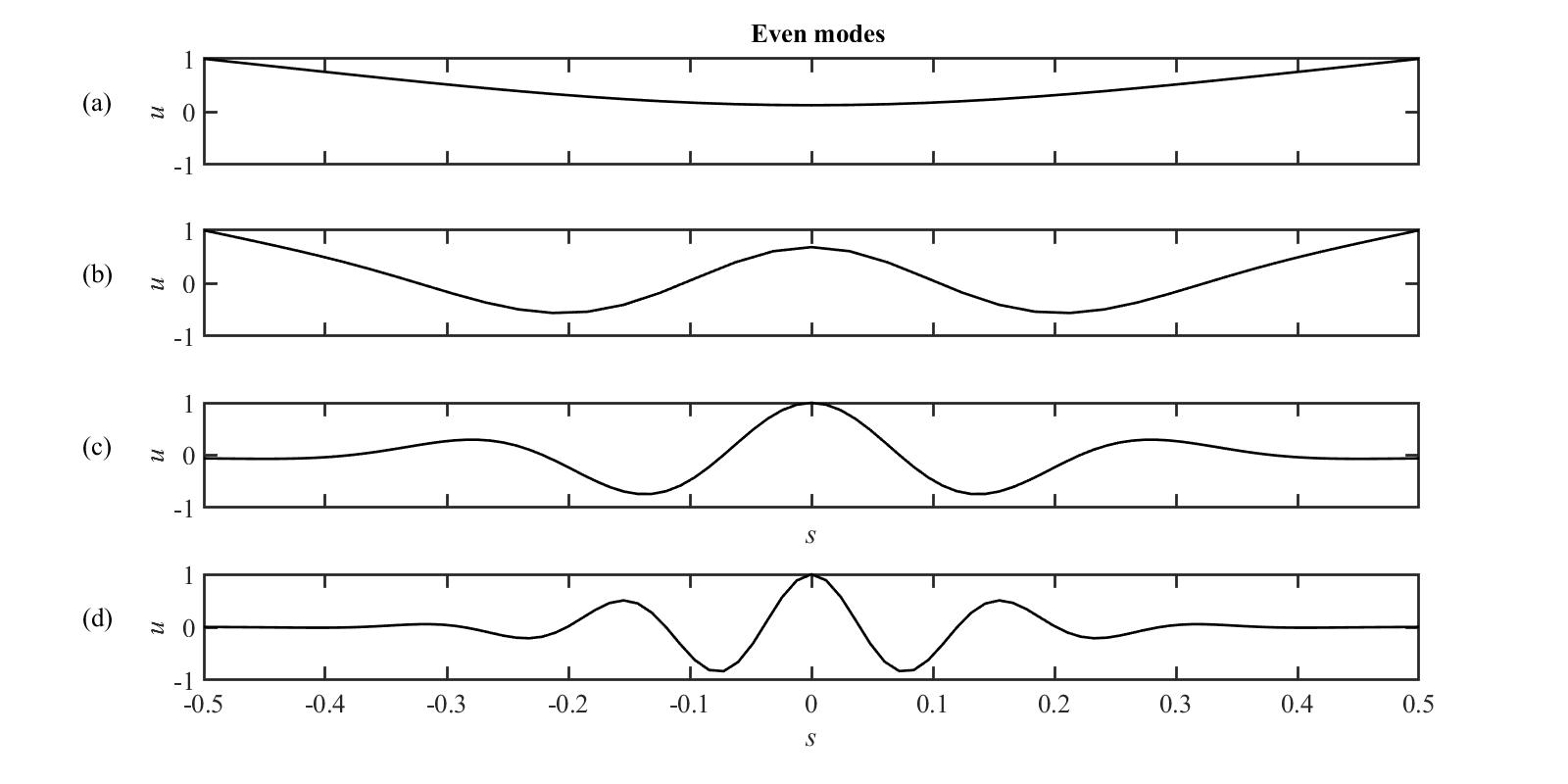}
	\vspace{-.85cm}
	\caption{
	Examples of the most unstable eigenfunctions $\Phi_u(s)$. Left, 
	 top-down:  $ -\eta \sin(2\phi)=3000$ (odd mode 2), 
	 10000 (odd mode 5), 
	 30000 and 
	 60000 (higher odd modes). Right, top-down: 
	  $-\eta \sin(2\phi)\!=\!300$ (even mode 1), $6000$ (even mode 4), $15000$ and $50000$ (higher even modes).
	 }
	\label{fig:4}
	\end{figure*}
	\begin{figure*}[http]
	\centerline{\includegraphics[width=0.7\textwidth]{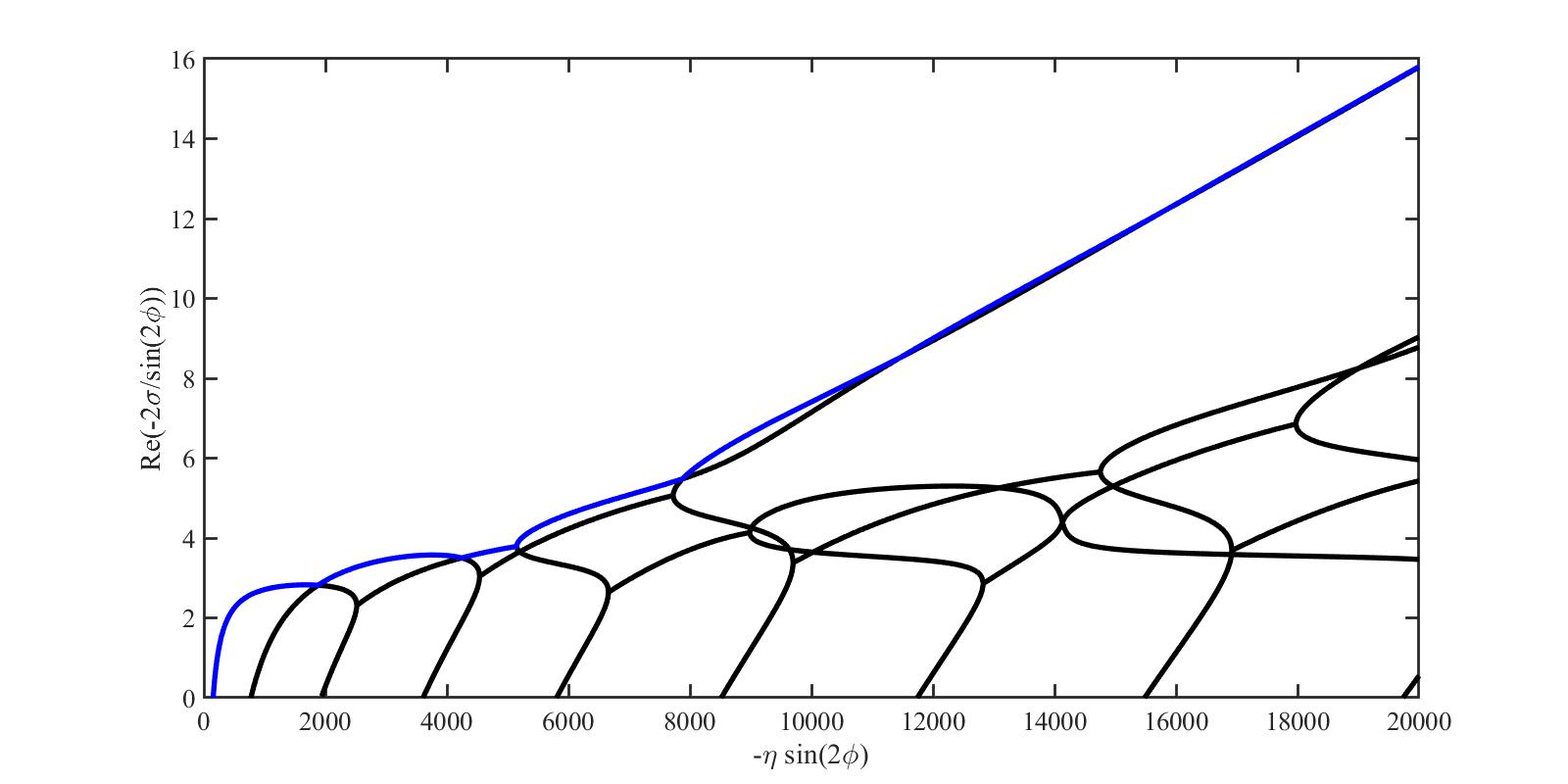}}
	\vspace{-0.4cm}
	\caption{Scaled eigenspectrum $\sigma=\sigma_u$ for perturbations $\Phi_u$ within the shear plane of the shear flow.}
	\label{fig:spectrum}
\end{figure*}

The modes 1-5 are the most unstable eigenfunctions $\Phi_u$ for the range of  the smallest positive eigenvalues $\sigma_u$, shown in our Fig.~\ref{fig:7} and Fig.~2 of Ref.~\cite{Becker2001}. The modes 1-5 appear for  $153.2\le -\eta \sin(2\phi)\le 10^4$. Their shapes are presented in the left panel of Fig.~\ref{fig:12}, using colors. 
The eigenfunctions $\Phi_u$ are normalized in such a way that $\max_s(\Phi_u(s))=1$. 
In the right panel of Fig.~\ref{fig:12}, shapes of the most unstable eigenfunctions $\Phi_u$ 
are shown 
for the whole range 
$0\le -\eta \sin(2\phi)\le 6\cdot 10^4$. 

The eigenfunctions $\Phi_u$ corresponding to the most unstable eigenvalues  are even for modes 1, 3, and 4, and odd for modes 2 and 5. In general, for certain values of $-\eta \sin(2\phi)$ they are odd, as in the left panel of Fig.~\ref{fig:4};  for other values they are even, as in the right panel of Fig.~\ref{fig:4}. 
At larger values of $-\eta \sin(2\phi)$, 
there appear 
\begin{figure*}[http]
	\centerline{\includegraphics[width=0.74\textwidth]{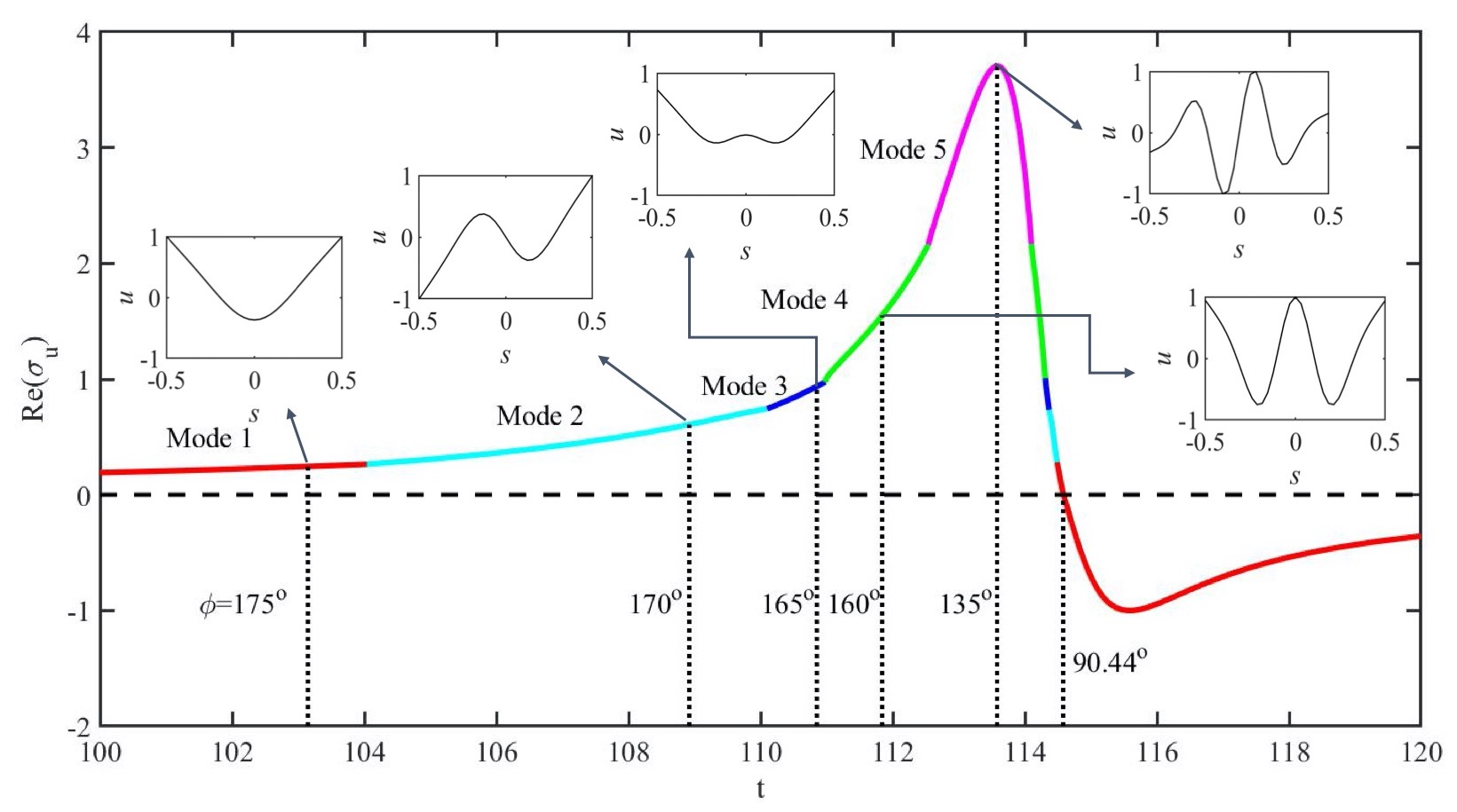}\hspace{.1cm}} 	\vspace{-0.4cm}
	\caption{Spectral problem for elastica evolving in time $t$ according to the zero-order Eqs~\eqref{eq:7}, assuming that elastica is straight with a very small perturbation only. The real part of the most unstable eigenvalue $\sigma_u$ 
	(in-plane perturbations) for different orientations $\bm{\lambda}$ of almost straight elastica with $\eta=10000$. Here $\theta=\pi/2$ and $\phi$ decreases with time from $\phi_0
	=179.5^{\circ}$ at $t=0$. 
The corresponding most unstable eigenfunctions are also shown. 
	}
	\label{fig:8}
\end{figure*}
buckling modes with larger wavenumbers. This finding seems to agree with more buckled shapes observed in the numerical simulations with the decreased bending stiffness, as shown in Fig. 12 of Ref. \cite{slowicka2022}.

For stability analysis, the most important are the largest eigenvalues. In the right panel of Fig. \ref{fig:2}, the largest eigenvalues for the in-plane and out-of-plane perturbations  are compared with each other. 
The eigenspectrum for Eqs \eqref{eq:12p} and \eqref{Lplanar}, i.e., for perturbations $\Phi_v$ perpendicular to  the shear plane, corresponds to  $\eta \sigma_v=\eta \sigma_u + \eta \sin(2\phi)/2$. Therefore, the  growth rate of the most unstable out-of-plane perturbations is smaller than the growth rate of the most unstable in-plane perturbations, as illustrated in the right panel of  Fig.~\ref{fig:7}. The zeros of the eigenvalues $\sigma_u$ and $\sigma_v$ correspond to $-\eta$sin$(2\phi)=\mu_1$ and $\mu_2$, respectively, with $\mu_1=153.2$ and $\mu_2=221.2$, as shown in the right panel of Fig. \ref{fig:7}. This means that the filament with an initial angle $\phi$ slightly smaller than $\pi$, while moving in the shear flow and decreasing the value of $\phi$, will first experience in-plane instability, and later out-of-plane instability. 

It is very interesting to point out that 
Eqs \eqref{eq:11p}-\eqref{Lplanar} are essentially the same as Eqs (3)-(4) in Ref. \cite{chakrabarti2020} for elastica in the compressional ambient flow $\bm{u}=(-x,y,0)$, with the following modifications: instead of $\bar{\mu}$, $\sigma$, $\Phi_y$ and $\Phi_z$ from Ref. \cite{chakrabarti2020}, here we have $-\eta \sin(2\phi)$, $-2\sigma/\sin(2\phi)$, $\Phi_u$ and $\Phi_v$, respectively.
Therefore, if we rescale our eigenvalues $\sigma_u$ by $-\frac{\sin(2\phi)}{2}$, a scaled eigenspectrum is obtained, shown in Fig.~\ref{fig:spectrum},
which is the same as that illustrated by Chakrabarti {\it{et al.}} \cite{chakrabarti2020} in their Fig. 3a for their unstable planar eigenspectrum in 
the compressional flow. 
The eigenspectra 
in shear and compressional flows 
are the same. 

When 
$-\eta \sin(2\phi)$ increases from 153.2 (critical value at which the buckling instability occurs) to 10$^4$, there appear 
even, odd, even, even, and odd modes, respectively. When $-\eta \sin(2\phi)$ increases above 10$^4$, 
the most unstable mode still keeps 
alternating odd and even symmetries.
This coupling of the even and odd most unstable modes was described and shown in Fig.~3 in Ref.~\cite{chakrabarti2020} in the context of the compressional flow. 

The spectral analysis provides 
the 
growth rate of small perturbations, given by Eq. \eqref{eks}, with the amplitudes independent of time. 
Keeping it in mind, in  Fig.~\ref{fig:8} we illustrate how the most unstable eigenfunctions and eigenvalues change when the filament is almost straight and its orientation $\bm{\lambda}$ is changed according to the zeroth order 
Eqs  \eqref{eq:7}-\eqref{eq:8}. 
In particular, as expected, 
the fastest growth of small perturbations is observed for $\phi\! =\!135^\circ$. The  unstable perturbations correspond only to angles $\phi$  larger than 
$90^{\circ}$ plus a certain small value. 
Fig.~\ref{fig:8} shows that the shapes and parity of the most unstable eigenfunctions are different for different orientations, for the same value of~$\eta$.

\section{Conclusions}
In this work, we derived Eqs \eqref{eq:9}-\eqref{eq:10} for the evolution of 3-dimensional perturbations of a straight elastica at an arbitrary orientation. 
We performed spectral analysis of elastica in a shear flow, investigating perturbations out of the shear plane. We analyzed the most unstable eigenfunctions and eigenvalues. We demonstrated that the eigenproblem for elastica in the shear flow is described by the same equations as in the compressional flow \cite{chakrabarti2020}.

Based on the spectral analysis presented here, in Ref.~\cite{slze} scaling law for the eigenfunction shape has been derived and compared with the numerical simulations of flexible filaments made of beads. In the future, it would be interesting to extend the spectral analysis presented here and in Ref.~\cite{slze} for an elastic sheet in a shear flow. Such a system has been recently studied, e.g., in Refs \cite{shelley2011,xu2014,yu2021,silmore2021,botto2022}, with potential applications for graphene flakes. 

\section*{Acknowledgements}
M. L. E.-J. thanks Prof. Howard A. Stone and  Dr.~Piotr Zdybel for helpful discussions. This work 
was supported in part by the National Science Centre under
grant UMO-2018/31/B/ST8/03640.

\bibliographystyle{ieeetr}
\bibliography{references}
\end{document}